\documentclass[%
superscriptaddress,
twocolumn,
showpacs,
preprintnumbers,
 amsmath,amssymb,
 aps,
floatfix,
]{revtex4-1}

\usepackage{graphicx}% Include figure files
\usepackage{dcolumn}% Align table columns on decimal point
\usepackage{bm}% bold math
\usepackage{hepunits}
\usepackage[dvipsnames]{xcolor}
\usepackage{textcomp}
\usepackage{lineno}
\usepackage{color}
\usepackage{float}
\begin{document}

%\preprint{APS/123-QED}

\title{Transition between Instability and Seeded Self-Modulation of a Relativistic Particle Bunch in Plasma}% Force line breaks with \\

\author{F. Batsch}
\affiliation{Max Planck Institute for Physics, Munich, Germany}
%\affiliation{CERN, Geneva, Switzerland}
%\affiliation{Technical University Munich, Munich, Germany}%Lines break automatically or can be forced with \\
\author{P. Muggli}%
\affiliation{Max Planck Institute for Physics, Munich, Germany}
\author{R.~Agnello}
\affiliation{Ecole Polytechnique Federale de Lausanne (EPFL), Swiss Plasma Center (SPC), Lausanne, Switzerland}
\author{C.C.~Ahdida}
\affiliation{CERN, Geneva, Switzerland}
\author{M.C.~Amoedo~Goncalves}
\affiliation{CERN, Geneva, Switzerland}
\author{Y.~Andrebe}
\affiliation{Ecole Polytechnique Federale de Lausanne (EPFL), Swiss Plasma Center (SPC), Lausanne, Switzerland}
\author{O.~Apsimon}
\affiliation{Cockcroft Institute, Daresbury, UK}
\affiliation{University of Liverpool, Liverpool, UK}
\author{R.~Apsimon}
\affiliation{Cockcroft Institute, Daresbury, UK}
\affiliation{Lancaster University, Lancaster, UK}
\author{A.-M.~Bachmann}
\affiliation{Max Planck Institute for Physics, Munich, Germany}
%\affiliation{CERN, Geneva, Switzerland}
%\affiliation{Technical University Munich, Munich, Germany}
\author{M.A.~Baistrukov}
\affiliation{Novosibirsk State University, Novosibirsk, Russia}
\affiliation{Budker Institute of Nuclear Physics SB RAS, Novosibirsk, Russia}
\author{P.~Blanchard}
\affiliation{Ecole Polytechnique Federale de Lausanne (EPFL), Swiss Plasma Center (SPC), Lausanne, Switzerland}
\author{F.~Braunm{\"u}ller}
\affiliation{Max Planck Institute for Physics, Munich, Germany}
\author{P.N.~Burrows}
\affiliation{John Adams Institute, Oxford University, Oxford, UK}
\author{B.~Buttensch{\"o}n}
\affiliation{Max Planck Institute for Plasma Physics, Greifswald, Germany}
\author{A.~Caldwell}
\affiliation{Max Planck Institute for Physics, Munich, Germany}
\author{J.~Chappell}
\affiliation{UCL, London, UK}
\author{E.~Chevallay}
\affiliation{CERN, Geneva, Switzerland}
\author{M.~Chung}
\affiliation{UNIST, Ulsan, Republic of Korea}
\author{D.A.~Cooke}
\affiliation{UCL, London, UK}
\author{H.~Damerau}
\affiliation{CERN, Geneva, Switzerland}
\author{C.~Davut}
\affiliation{Cockcroft Institute, Daresbury, UK}
\affiliation{University of Manchester, Manchester, UK}
\author{G.~Demeter}
\affiliation{Wigner Research Center for Physics, Budapest, Hungary}
\author{H.L.~Deubner}
\affiliation{Philipps-Universit{\"a}t Marburg, Marburg, Germany}
%\author{A.~Dexter}
%\affiliation{Cockcroft Institute, Daresbury, UK}
%\affiliation{Lancaster University, Lancaster, UK}
\author{S.~Doebert}
\affiliation{CERN, Geneva, Switzerland}
\author{J.~Farmer}
\affiliation{Max Planck Institute for Physics, Munich, Germany}
\affiliation{CERN, Geneva, Switzerland}
\author{A.~Fasoli}
\affiliation{Ecole Polytechnique Federale de Lausanne (EPFL), Swiss Plasma Center (SPC), Lausanne, Switzerland}
\author{V.N.~Fedosseev}
\affiliation{CERN, Geneva, Switzerland}
\author{R.~Fiorito}
\affiliation{Cockcroft Institute, Daresbury, UK}
\affiliation{University of Liverpool, Liverpool, UK}
\author{R.A.~Fonseca}
\affiliation{ISCTE - Instituto Universit\'{e}ario de Lisboa, Portugal}
\affiliation{GoLP/Instituto de Plasmas e Fus\~{a}o Nuclear, Instituto Superior T\'{e}cnico, Universidade de Lisboa, Lisbon, Portugal}
\author{F.~Friebel}
\affiliation{CERN, Geneva, Switzerland}
\author{I.~Furno}
\affiliation{Ecole Polytechnique Federale de Lausanne (EPFL), Swiss Plasma Center (SPC), Lausanne, Switzerland}
\author{L.~Garolfi}
\affiliation{TRIUMF, Vancouver, Canada}
\author{S.~Gessner}
\affiliation{CERN, Geneva, Switzerland} 
\affiliation{SLAC National Accelerator Laboratory, Menlo Park, CA, USA} 
\author{I.~Gorgisyan}
\affiliation{CERN, Geneva, Switzerland}
\author{A.A.~Gorn}
\affiliation{Novosibirsk State University, Novosibirsk, Russia}
\affiliation{Budker Institute of Nuclear Physics SB RAS, Novosibirsk, Russia}
\author{E.~Granados}
\affiliation{CERN, Geneva, Switzerland}
\author{M.~Granetzny}
\affiliation{University of Wisconsin, Madison, Wisconsin, USA}
\author{T.~Graubner}
\affiliation{Philipps-Universit{\"a}t Marburg, Marburg, Germany}
\author{O.~Grulke}
\affiliation{Max Planck Institute for Plasma Physics, Greifswald, Germany}
\affiliation{Technical University of Denmark, Lyngby, Denmark}
\author{E.~Gschwendtner}
\affiliation{CERN, Geneva, Switzerland} 
\author{V.~Hafych}
\affiliation{Max Planck Institute for Physics, Munich, Germany}
\author{A.~Helm}
\affiliation{GoLP/Instituto de Plasmas e Fus\~{a}o Nuclear, Instituto Superior T\'{e}cnico, Universidade de Lisboa, Lisbon, Portugal}
\author{J.R.~Henderson}
\affiliation{Cockcroft Institute, Daresbury, UK}
\affiliation{Accelerator Science and Technology Centre, ASTeC, STFC Daresbury Laboratory, Warrington, UK}
\author{M.~H{\"u}ther}
\affiliation{Max Planck Institute for Physics, Munich, Germany}
%\affiliation{Technical University Munich, Munich, Germany}
\author{I.Yu.~Kargapolov}
\affiliation{Novosibirsk State University, Novosibirsk, Russia}
\affiliation{Budker Institute of Nuclear Physics SB RAS, Novosibirsk, Russia}
\author{S.-Y.~Kim}
\affiliation{UNIST, Ulsan, Republic of Korea}
\author{F.~Kraus}
\affiliation{Philipps-Universit{\"a}t Marburg, Marburg, Germany}
\author{M.~Krupa}
\affiliation{CERN, Geneva, Switzerland}
\author{T.~Lefevre}
\affiliation{CERN, Geneva, Switzerland}
\author{L.~Liang}
\affiliation{Cockcroft Institute, Daresbury, UK}
\affiliation{University of Manchester, Manchester, UK}
\author{S.~Liu}
\affiliation{TRIUMF, Vancouver, Canada}
\author{N.~Lopes}
\affiliation{GoLP/Instituto de Plasmas e Fus\~{a}o Nuclear, Instituto Superior T\'{e}cnico, Universidade de Lisboa, Lisbon, Portugal}
\author{K.V.~Lotov}
\affiliation{Novosibirsk State University, Novosibirsk, Russia}
\affiliation{Budker Institute of Nuclear Physics SB RAS, Novosibirsk, Russia}
\author{M.~Martyanov}
\affiliation{Max Planck Institute for Physics, Munich, Germany}
\author{S.~Mazzoni}
\affiliation{CERN, Geneva, Switzerland}
\author{D.~Medina~Godoy}
\affiliation{CERN, Geneva, Switzerland}
\author{V.A.~Minakov}
\affiliation{Novosibirsk State University, Novosibirsk, Russia}
\affiliation{Budker Institute of Nuclear Physics SB RAS, Novosibirsk, Russia}
\author{J.T.~Moody}
\affiliation{Max Planck Institute for Physics, Munich, Germany}
\author{K.~Moon}
\affiliation{UNIST, Ulsan, Republic of Korea}
\author{P.I.~Morales~Guzm\'{a}n}
\affiliation{Max Planck Institute for Physics, Munich, Germany}
%\affiliation{Technical University Munich, Munich, Germany}
\author{M.~Moreira}
\affiliation{CERN, Geneva, Switzerland}
\affiliation{GoLP/Instituto de Plasmas e Fus\~{a}o Nuclear, Instituto Superior T\'{e}cnico, Universidade de Lisboa, Lisbon, Portugal}
\author{T.~Nechaeva}
\affiliation{Max Planck Institute for Physics, Munich, Germany}
\author{E.~Nowak}
\affiliation{CERN, Geneva, Switzerland}
\author{C.~Pakuza}
\affiliation{John Adams Institute, Oxford University, Oxford, UK}
\author{H.~Panuganti}
\affiliation{CERN, Geneva, Switzerland} 
\author{A.~Pardons}
\affiliation{CERN, Geneva, Switzerland}
\author{A.~Perera}
\affiliation{Cockcroft Institute, Daresbury, UK}
\affiliation{University of Liverpool, Liverpool, UK}
\author{J.~Pucek}
\affiliation{Max Planck Institute for Physics, Munich, Germany}
%\affiliation{Technical University Munich, Munich, Germany}
\author{A.~Pukhov}
\affiliation{Heinrich-Heine-Universit{\"a}t D{\"u}sseldorf, D{\"u}sseldorf, Germany}
\author{R.L.~Ramjiawan}
\affiliation{CERN, Geneva, Switzerland}
\affiliation{John Adams Institute, Oxford University, Oxford, UK}
\author{S.~Rey}
\affiliation{CERN, Geneva, Switzerland}
\author{K.~Rieger}
\affiliation{Max Planck Institute for Physics, Munich, Germany}
\author{O.~Schmitz}
\affiliation{University of Wisconsin, Madison, Wisconsin, USA}
\author{E.~Senes}
\affiliation{CERN, Geneva, Switzerland}
\affiliation{John Adams Institute, Oxford University, Oxford, UK}
\author{L.O.~Silva}
\affiliation{GoLP/Instituto de Plasmas e Fus\~{a}o Nuclear, Instituto Superior T\'{e}cnico, Universidade de Lisboa, Lisbon, Portugal}
\author{R.~Speroni}
\affiliation{CERN, Geneva, Switzerland}
\author{R.I.~Spitsyn}
\affiliation{Novosibirsk State University, Novosibirsk, Russia}
\affiliation{Budker Institute of Nuclear Physics SB RAS, Novosibirsk, Russia}
\author{C.~Stollberg}
\affiliation{Ecole Polytechnique Federale de Lausanne (EPFL), Swiss Plasma Center (SPC), Lausanne, Switzerland}
\author{A.~Sublet}
\affiliation{CERN, Geneva, Switzerland}
\author{A.~Topaloudis}
\affiliation{CERN, Geneva, Switzerland}
\author{N.~Torrado}
\affiliation{GoLP/Instituto de Plasmas e Fus\~{a}o Nuclear, Instituto Superior T\'{e}cnico, Universidade de Lisboa, Lisbon, Portugal}
\author{P.V.~Tuev}
\affiliation{Novosibirsk State University, Novosibirsk, Russia}
\affiliation{Budker Institute of Nuclear Physics SB RAS, Novosibirsk, Russia}
\author{M.~Turner}
\affiliation{CERN, Geneva, Switzerland}
\affiliation{Lawrence Berkeley National Laboratory, Berkeley, CA, USA}
\author{F.~Velotti}
\affiliation{CERN, Geneva, Switzerland}
\author{L.~Verra}
\affiliation{Max Planck Institute for Physics, Munich, Germany}
\affiliation{CERN, Geneva, Switzerland}
\affiliation{Technical University Munich, Munich, Germany}
\author{V.A.~Verzilov}
\affiliation{TRIUMF, Vancouver, Canada} 
\author{J.~Vieira}
\affiliation{GoLP/Instituto de Plasmas e Fus\~{a}o Nuclear, Instituto Superior T\'{e}cnico, Universidade de Lisboa, Lisbon, Portugal}
\author{H.~Vincke}
\affiliation{CERN, Geneva, Switzerland}
\author{C.P.~Welsch}
\affiliation{Cockcroft Institute, Daresbury, UK}
\affiliation{University of Liverpool, Liverpool, UK}
\author{M.~Wendt}
\affiliation{CERN, Geneva, Switzerland}
\author{M.~Wing}
\affiliation{UCL, London, UK}
\author{P.~Wiwattananon}
\affiliation{CERN, Geneva, Switzerland}
\author{J.~Wolfenden}
\affiliation{Cockcroft Institute, Daresbury, UK}
\affiliation{University of Liverpool, Liverpool, UK}
\author{B.~Woolley}
\affiliation{CERN, Geneva, Switzerland}
\author{G.~Xia}
\affiliation{Cockcroft Institute, Daresbury, UK}
\affiliation{University of Manchester, Manchester, UK}
\author{M.~Zepp}
\affiliation{University of Wisconsin, Madison, Wisconsin, USA}
\author{G.~Zevi~Della~Porta}
\affiliation{CERN, Geneva, Switzerland}
\collaboration{The AWAKE Collaboration}
\noaffiliation

%\affiliation{
%\affiliation{

\date{\today}% It is always \today, today,

\begin{abstract}
%abstract < 600 characters, including spaces
%%%
We use a relativistic ionization front to provide various initial transverse wakefield amplitudes for the self-modulation of a long proton bunch in plasma. %
We show experimentally that, with sufficient initial amplitude ($\ge(4.1\pm0.4)$\,MV/m), the phase of the modulation along the bunch is reproducible from event to event, with 3 to 7\% (of 2$\pi$) rms variations all along the bunch. %
The phase is not reproducible for lower initial amplitudes. %
We observe the transition between these two regimes. %
Phase reproducibility is essential for deterministic external injection of particles to be accelerated. %~565 characters with spaces
%\begin{description}
%item[PACS numbers]
%\end{description}
\end{abstract}

\pacs{Valid PACS appear here}% PACS, the Physics and Astronomy
\maketitle

%\tableofcontents
%\linenumbers\relax
\section{\label{sec:level1}Introduction}
Accelerators rely on precise control of parameters to produce high-quality, high-energy particle bunches for numerous applications. %
A class of novel accelerators using plasma as a medium to sustain large accelerating~\cite{bib:tajima,bib:chenpwfa} and focusing~\cite{bib:chenfoc} fields has emerged and has made remarkable experimental progress over the last two decades~\cite{bib:blumenfeld,bib:mike,bib:benedeti}. %

Most of these accelerators use a very short ($<$1\,ps), intense laser pulse~\cite{bib:tajima} or a dense, relativistic particle bunch~\cite{bib:chenpwfa} to drive wakefields in plasma. %
The amplitude of the accelerating field that can be sustained with a plasma of electron density $n_{e0}$ is on the order of the wave breaking field~\cite{bib:wavebreaking}: $E_{WB}=\left(m_ec/e\right)\omega_{pe}$. %
Here $\omega_{pe}=\left(n_{e0}e^2/\varepsilon_0 m_e\right)^{1/2}$ is the plasma electron angular frequency~\cite{bib:cst}. %
Assuming the driver of rms duration $\sigma_{t}$ fits within the structure, i.e., $\sigma_{t}\cong1/\omega_{pe}$, one can re-write: %$E_{WB}< \frac{m_ec^2}{e}\frac{1}{\sigma_{t}}$
$E_{WB}=\left(m_ec/e\right)\left(1/\sigma_{t}\right)$. %
Therefore operating at high accelerating field ($>$1\,GV/m) requires high plasma density and short ($<$2\,ps) pulses or bunches with similarly small radii ($\sigma_{r0}\le c/\omega_{pe}\le$600\,\textmu m) \cite{bib:su}. %

The system extracts energy from the driver and transfers it to a witness bunch, through the plasma. %
The total energy gain of the witness bunch is limited to the energy carried by the driver. %
Short laser pulses and particle bunches available today and suitable to drive $>$1\,GV/m amplitude wakefields carry less than $\sim$100\,J of energy. %
Laser pulses and particle bunches carrying much more energy are too long, typically $>$100\,ps, to drive large amplitude wakefields when following the above $E_{WB}\propto1/\sigma_{t}$ scaling. %
%However, a long laser pulse~\cite{bib:laserSM} or a long, relativistic particle bunch~\cite{bib:protonSM} propagating in dense plasma, i.e., $\sigma_{t}\gg1/\omega_{pe}$, is subject to instabilities that, through self-modulation (SM), can transform it into a train of pulses/bunches shorter than, and separated by $2\pi/\omega_{pe}$. The train can then resonantly excite large amplitude wakefields. %
%
However, long laser pulses~\cite{bib:laserSM} and long, relativistic particle bunches~\cite{bib:protonSM} propagating in dense plasma, i.e., $\sigma_{t}\gg1/\omega_{pe}$, are subject to self-modulation (SM) instabilities. % that, through self-modulation (SM),
These instabilities can transform them into a train of pulses/bunches shorter than, and with a periodicity of  $2\pi/\omega_{pe}$. The train can then resonantly excite large amplitude wakefields. %
Control of the SM process, in particular of the relative phase of the wakefields, is necessary to deterministically inject a witness bunch shorter than $1/\omega_{pe}$ into the accelerating \emph{and} focusing phase of the wakefields. %

As the first proton-driven plasma wakefield acceleration (PWFA)  experiment, AWAKE~\cite{bib:awake,bib:muggli} recently demonstrated that the SM process does indeed transform a long proton bunch ($\sigma_t>200$\,ps) into a train of micro-bunches with period $2\pi/\omega_{pe}$ ($<10$\,ps)~\cite{bib:karl}. %
We also demonstrated that the process grows along the bunch \emph{and} along the plasma, from the initial wakefield amplitude, to saturate at much larger values~\cite{bib:marlene,bib:marlene2}. %
Electrons were externally injected in the wakefields, though without phase control (electron bunch duration on the order of $2\pi/\omega_{pe}$) and accelerated from $\sim$19\,MeV to $\sim$2\,GeV~\cite{bib:awakeaccel}. %
For this scheme to become an accelerator that can produce not only sufficiently high-energy particles, but also sufficiently high-quality bunches in terms of high population, low energy spread and low emittance~\cite{bib:veronica}, one needs to show that the SM process can be controlled. %

Seeding of SM in the sense of triggering the start of its growth has been demonstrated experimentally with a relativistic ionization front (RIF) in a long pulse, laser-driven wakefield accelerator~\cite{bib:leblanc}, and with the sharp density front of a long electron bunch in a PWFA~\cite{bib:fang}. %
%However, no measurements of the effect of that seeding on the phase of the wakefields were performed. %
However, measurements on the effect of that seeding on the phase of growing wakefields have not been reported.
As demonstrated below, triggering the SM is not sufficient to ensure the reproducibility of the phase of the wakefields from event to event. %

In this \emph{Letter}, we demonstrate experimentally for the first time that the SM of a long, relativistic particle bunch can be seeded by a RIF. %
We define seeding as the conditions leading to a reproducible timing/phase of the SM along the bunch with respect to the RIF. %
From time-resolved images of the bunch obtained at two plasma densities, we analyze the relative timing/phase of the micro-bunches along the proton bunch, after the plasma. We control the initial wakefield amplitude through the timing of the RIF along the bunch. %
When the process is not seeded, we observe randomly distributed phases and thus the SM instability (SMI)~\cite{bib:protonSM}. %
With sufficiently strong initial amplitude, the phase of the wakefields varies by only a small fraction of 2$\pi$ from event to event, the characteristic of seeded SM (SSM)~\cite{bib:muggli}. This is despite natural variations of the incoming bunch parameters~\cite{bib:moreira}. 
We thus observe the transition from SMI to SSM. %
We also observe phase reproducibility over more than 2$\sigma_{t}$ along the bunch. %
Phase reproducibility is essential for future experiments~\cite{bib:muggli} with deterministic, external injection of particles to be accelerated (e$^-$ or e$^+$) at a precise phase within the accelerating and focusing region of the wakefields~\cite{bib:veronica}. %

%Besides this possible application to a future accelerator, the experimental results presented here show that the phase of the self-modulation instability can be controlled and the process seeded, an interesting and significant physics result. %
Experimental results presented here show that the phase of the self-modulation instability, a fundamental beam/plasma interaction mechanism~\cite{bib:protonSM}, can be controlled. %
It is also a requirement for future acceleration experiments. %

\section{\label{sec:level2}Experimental Setup}%
\begin{figure}[b]
\centering
\includegraphics[width=\columnwidth]{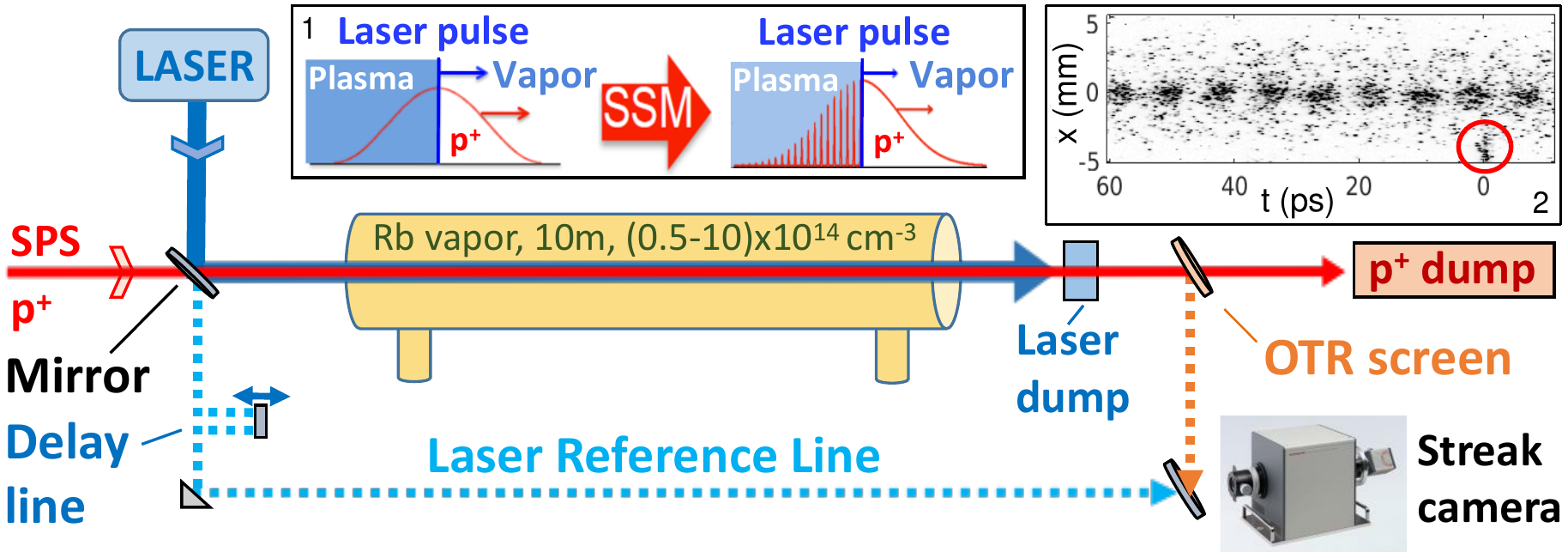}
\caption{Schematic of the experimental setup showing the main components used for measurements presented here. Inset 1: RIF in the middle of the proton bunch ($t_{RIF}=0$\,ps). % 
Inset 2: Streak camera image of a modulated proton bunch, laser reference signal at $t=0$\,ps (red circle).}
\label{fig:setup}
\end{figure}

The CERN Super Proton Synchrotron (SPS) provides a Gaussian bunch with 400\,GeV energy per proton, 3$\times$10$^{11}$ particles and a rms duration $\sigma_{t}=250$\,ps. 
The bunch enters a 10 m-long vapor source \cite{bib:gennady,bib:erdem}, as shown in Fig.~\ref{fig:setup}, with rms waist size $\sigma_{r0}$=150\,\textmu m.
The source contains rubidium (Rb) vapor with density $n_{Rb}$ adjustable in the  (0.5~-~10)$~\times$~10$^{14}$\,cm$^{-3}$ range and with uniform temperature and thus density distributions ($\frac{\Delta n_{Rb}}{n_{Rb}}=\frac{\Delta T}{T}<$0.2\% \cite{ bib:erdem}). %
The vapor density is measured to better than 0.5\% \cite{bib:fabian} at both ends of the source. 
A Titanium:Sapphire laser system provides a 120\,fs, $\le$450\,mJ laser pulse that can serve two purposes. % 
First, when propagating along the vapor column it creates the plasma at the RIF. %
The RIF transforms the Rb vapor into a $\sim$~2\,mm diameter plasma with density and uniformity equal to those of the vapor~\cite{bib:karl}. %
Therefore, hereafter we quote the corresponding plasma density instead of the measured Rb vapor density ($n_{e0}$=$n_{Rb})$.
Second, when propagating within the proton bunch, the RIF triggers the sudden ($\ll 1/\omega_{pe}$) onset of beam plasma interaction that can seed the SM process. %
Seeding can occur because this onset corresponds to the driving of initial plasma wakefields starting at the RIF and with amplitudes depending on the local bunch density~\cite{bib:karl,bib:marlene}. %

The train of micro-bunches resulting from the SM process leaves the plasma after 10\,m and passes through an aluminum-coated screen where protons emit optical transition radiation (OTR), 3.5\,m from the plasma exit. %
The OTR has the same spatio-temporal structure as the modulated proton bunch. %
A streak camera resolves the incoming OTR light imaged onto its entrance slit in space and in time with resolutions of 80\,\textmu m and $\sim$1\,ps, respectively, over a 73\,ps time window. %
Since the entrance slit is narrower than the bunch radius at the screen location, images display the bunch charge density and not its charge~\cite{bib:annamaria}. %
A transfer line (dashed blue line in Fig.~\ref{fig:setup}, \cite{bib:fabianEAAC}) guides a mirror bleed-through of the laser pulse to the streak camera. %
This signal (in red circle in Inset\,2 of Fig.~\ref{fig:setup}) indicates on each image the relative timing of the RIF within the proton bunch with 0.53\,ps (rms) accuracy and 0.16\,ps precision. It can be delayed together with the camera trigger signal to appear on the image at times later than that of the RIF, as seen every 50\,ps at the bottom of Fig.~\ref{fig:alongbunch}~(a). %
This signal is necessary to refer images in time with respect to the RIF’s and with respect to each other’s timing, because the streak camera triggering system has a time jitter of 4.8\,ps (rms), equivalent to approximately half a period of the wakefields. % 
In the following, we refer to this signal as the laser reference signal (LRS). % 
  
%%Physics starts here ...
\section{\label{sec:level3}Results}
We observe that when we use the RIF for plasma creation only, placing it nano- to micro-seconds ahead of the proton bunch, SM occurs~\cite{bib:gessner}. %
In this case SM can grow from noise present in the system. %
The wakefield amplitude driven by shot noise in the proton bunch distribution was estimated at the tens of kV/m level~\cite{bib:lotovnoise}. %
The laser pulse drives wakefields at the $<$100\,kV/m level at the plasma densities of these experiments~\cite{bib:muggliEAAC}. % 
\begin{figure}[bth]
\centering
\includegraphics[width=\columnwidth]{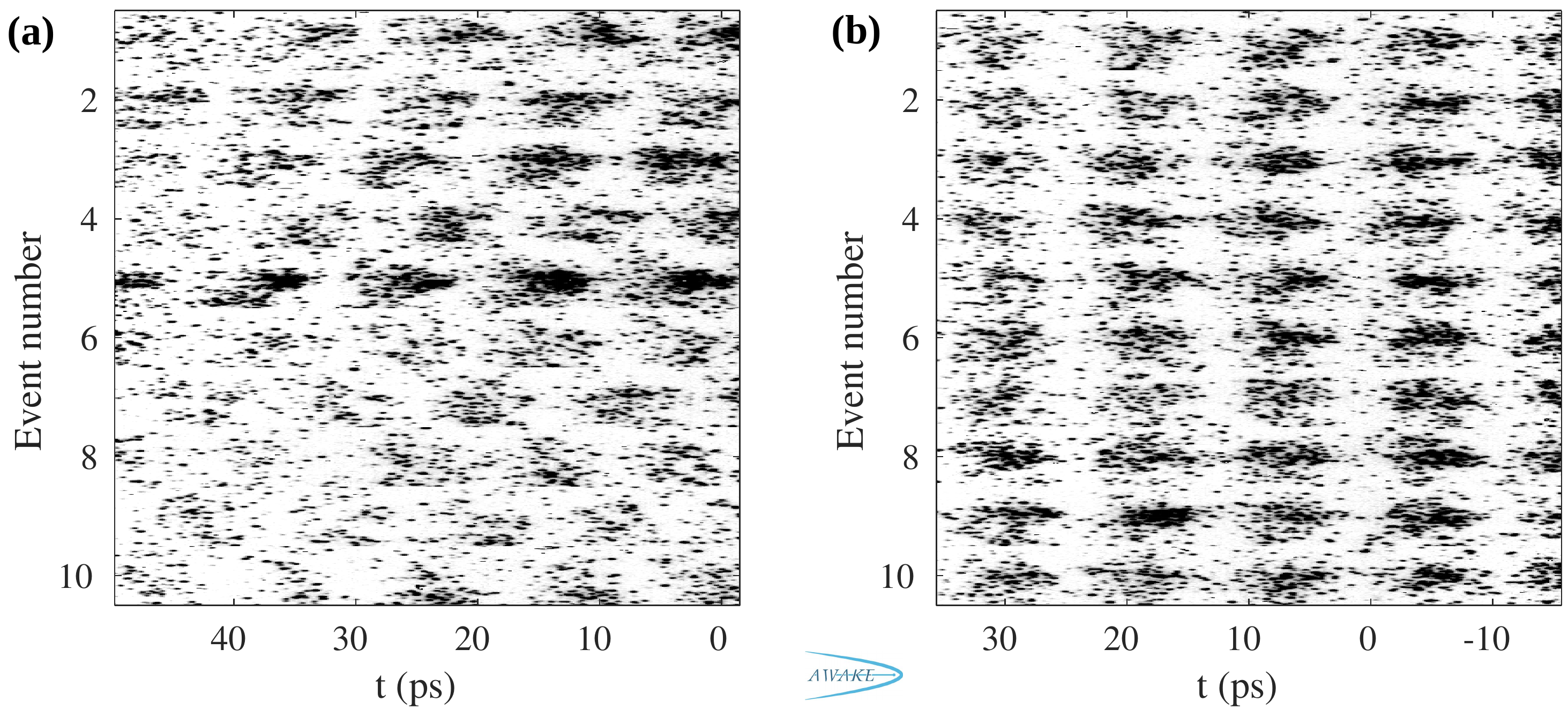}
\caption{Composite images of the center part of the streak camera image (see Fig.~\ref{fig:setup} Inset2) for ten events with (a) RIF 600\,ps (2.4$\sigma_t$) and (b) RIF 350\,ps (1.4$\sigma_t$) ahead of the proton bunch center. Front of the bunch on the right hand side. %
Events aligned with respect to LRS (\cite{bib:fabianEAAC}, at $t=0$, not visible). %
Both cases: LRS 150\,ps (0.6$\sigma_t$) ahead of bunch center, $n_{e0}=0.94\times$10$^{14}$\,cm$^{-3}$. %Time scales are different because of variations of the LRS's timing in the 73\,ps acquisition windows.
}
\label{fig:waterfall}
\end{figure}
Figure~\ref{fig:waterfall}~(a) shows a composite image of the time structure of the center part of the modulated proton bunch (compare Fig.~\ref{fig:setup} Inset 2) for ten events in the 73\,ps window, placed 150\,ps (0.6$\sigma_{t}$) ahead of the bunch peak. %
These events are aligned in time with respect to the LRS. %
%A plasma density of ($n_{e0}=0.94\times10^{14}\,\mathrm{cm}^{-3}$) was chosen to provide a sufficient resolution of the modulation profile.
The LRS alignment procedure yields a $\sim$50\,ps-long common window between images. %
The LRS (not shown) is placed at $t$=0\,ps on each image. %
The RIF is 600\,ps (2.4$\sigma_{t}$) ahead of the bunch peak (i.e., 450\,ps, 1.8$\sigma_{t}$ between RIF and $t$=0 on the image). %
Each image is normalized to its incoming bunch population. %
The figure clearly shows that from event to event micro-bunches appear at no particular times with respect to the RIF. %
It also shows that the measured micro-bunch charge density varies considerably. Variations in bunch density on these images can be attributed to amplitude variations of focusing and defocusing fields~\cite{bib:annamaria}. %
Variations in timing/phase and amplitude of the modulation are expected for the occurrence of a (non-seeded) instability such as SMI~\cite{bib:protonSM}. %

Figure~\ref{fig:waterfall}~(b) shows a similar plot to that of Fig.~\ref{fig:waterfall}~(a), but with the RIF placed closer, 350\,ps (1.4$\sigma_t$) ahead of the bunch peak and thus with larger wakefield amplitude at the RIF, with all other parameters unchanged. % 
It is clear that in this case the micro-bunches appear essentially at the same time with respect to the RIF  and with much more consistent charge density than in the previous case. %
This data shows the behavior expected from a seeded process such as SSM. %
From these two plots we conclude that in the first case the phase of the modulation is not reproducible from event to event (SMI), whereas it is in the second case (SSM). %

\begin{figure*}[bthp!]
\centering
\includegraphics[width=\textwidth]{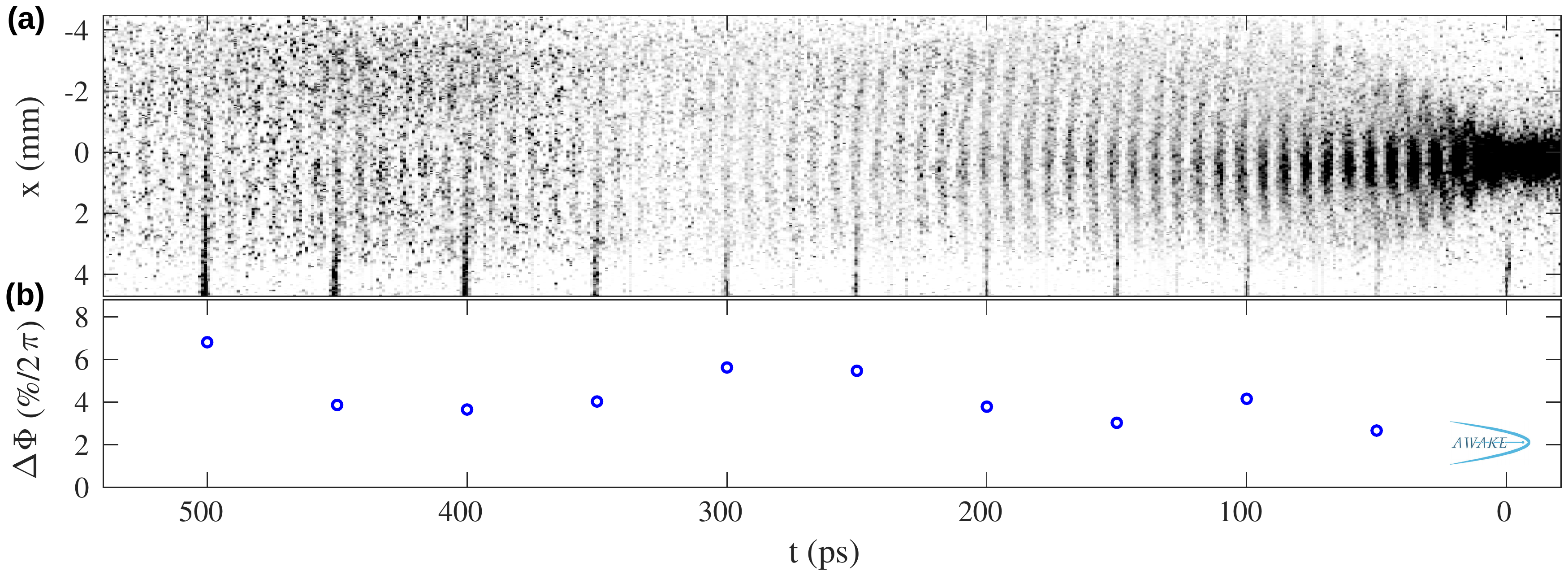}
\caption{(a) Time-resolved, ``stitched" image of the self-modulated proton bunch with $t_{RIF}=125$\,ps (0.5$\sigma_t$), $n_{e0}=1.81\times$10$^{14}$\,cm$^{-3}$. The RIF is at $t$=0 on the image (not visible). The LRS is visible every 50\,ps at the bottom of the image. (b) Modulation rms phase variation for each set of images with equal LRS timing.}
\label{fig:alongbunch}
\end{figure*}

In order to quantify the observed effect, we determine the phase/timing (using the modulation frequency/period) of the bunch modulation with respect to the RIF. %
For this purpose we sum counts of the bunch image in a $\cong\pm$430\,\textmu m-wide region around the axis of the bunch at the OTR screen to %
obtain a time profile of the bunch SM. % (see Supplemental Material). %
At this location the incoming bunch transverse rms size is $\cong574$\,\textmu m (see Fig.~\ref{fig:alongbunch} (a), $t$$<$0\,ps). %
For each event, we determine the time of the LRS in the 73\,ps window. %
We calculate the relative phase/timing of the micro-bunch appearing after the LRS as explained in the \emph{Supplemental Material}. %
For the data set analyzed here ($n_{e0}=0.94\times10^{14}\,\mathrm{cm}^{-3}$), the modulation frequency is 87.1\,GHz. %

\begin{figure}[hbt]
\centering
\includegraphics[width=\columnwidth]{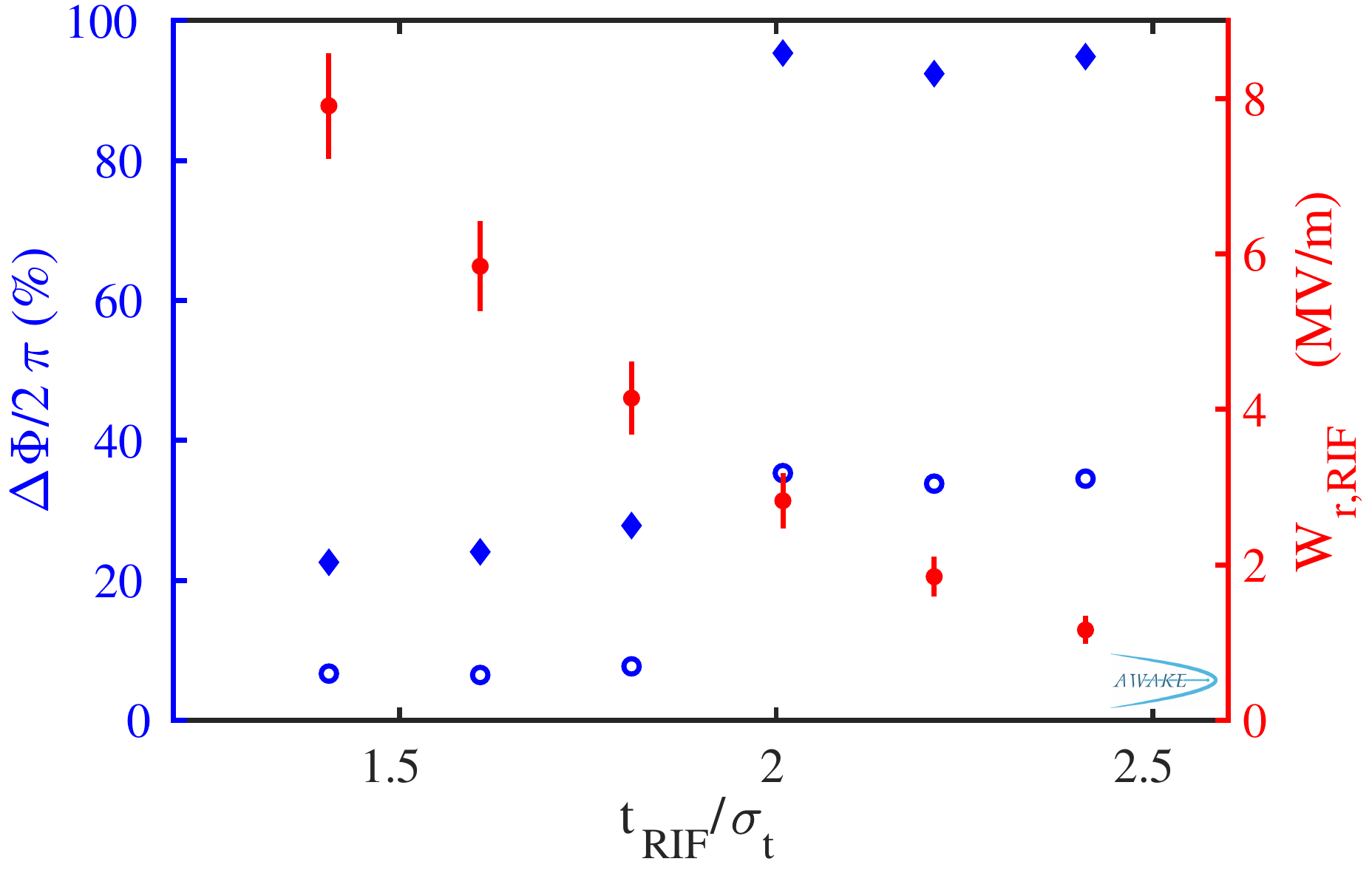}
\caption{Measured rms (blue circles) and full phase variation (blue diamond), and initial linear transverse wakefield amplitude (filled red circles) as a function of $t_{RIF}$ normalized to $\sigma_t$. %For $t_{RIF}\leq1.8\sigma_{t}$, the circles show the rms value of the relative phase variations. 
The error bars indicate the statistical uncertainty of 10.1\% (see text). Error bars representing the uncertainty in $t_{RIF}$ due to the 15\,ps (0.06$\sigma_t$) rms proton timing jitter are not plotted. %
Same LRS timing and $n_{e0}$ as in Fig.~\ref{fig:waterfall}.}
\label{fig:transitionplot}
\end{figure}

Figure~\ref{fig:transitionplot} shows the variation in relative phase for six series (including the events of Fig.~\ref{fig:waterfall}) of approximately 18 events each, measured with the analysis window (and LRS) 150\,ps ahead of the bunch peak, as a function of the RIF timing $t_{RIF}$ along the bunch normalized to the rms bunch duration. %
%Quoting a phase variation value is meaningful only when this variation is a small fraction of its full possible range of (modulo) 2$\pi$. % (or multiples of it). %
%Therefore, in Fig.~\ref{fig:transitionplot} we plot the rms value of the relative phase variations only when it is $\ll2\pi$ (blue circles). %
%In all cases we plot the full range of values observed (blue diamonds). %
%in Fig.~\ref{fig:transitionplot} we plot the rms value of the relative phase variations (blue circles) and the full range of values observed (blue diamonds).
%Since the number of events recorded is limited, the measured phase range does not necessarily cover a full $2\pi$  in the SMI case, but a large fraction of it. %
%This plot shows a clear transition in phase variations between large ones ($\sim2\pi$ range, $\sim$35\% rms) for $t_{RIF}\ge$2.0$\sigma_{t}$ and small ones ($\ll2\pi$ range, $\sim$6\% rms) for $t_{RIF}\le$1.8$\sigma_{t}$. %
%
%The phase distributions for $t_{RIF}\le$1.8$\sigma_{t}$ cover a range (blue diamonds) $\ll$2$\pi$ and their rms (blue circles) is small, $\sim$6\%, which shows that the phase of the SM is reproducible from event to event (within the rms range). %
%On the contrary, for  $t_{RIF}\ge$2.0$\sigma_{t}$, the ranges cover close to 2$\pi$ and their rms approaches the value expected for a uniform distribution, 29\%. % 
%This corresponds to a phase randomly distributed from event to event, possibly varying over more than 2$\pi$. %
The phase distributions for $t_{RIF}\ge$2.0$\sigma_{t}$ cover a range (blue diamonds) close to 2$\pi$ and their rms (blue circles) approaches  the value expected for a uniform distribution, 29\%. % 
This corresponds to a phase randomly distributed from event to event, possibly varying over more than 2$\pi$. %
On the contrary, for $t_{RIF}\le$1.8$\sigma_{t}$, the ranges are $\ll$2$\pi$ and their rms is small, $\sim$6\%, which shows that the phase of the SM is reproducible from event to event (within the rms range). %
This is the transition from SMI, with the modulation phase not reproducible (Fig.~\ref{fig:waterfall}~(a)), to SSM, with the modulation phase reproducible within a small range of 2$\pi$ (Fig.~\ref{fig:waterfall}~(b)), when the initial wakefield amplitude increases. %
We show later, by delaying the observation window timing for a fixed $t_{RIF}$, that when reached in one window, the timing/phase reproducibility occurs all along the bunch, as expected. %Move???? after
In the SMI regime, time-resolved images (not presented here) of the SM near the seed point show that full SM starts at different times along the bunch, unlike in the seeded cases, where it starts at the RIF \cite{bib:karl}. %,bib:leblanc}. %
This explains the $\sim$2$\pi$ (modulo) phase variations observed with SMI. In the SSM regime, the observed phase rms variations of $\sim$\,6\% (of $2\pi$) results from  at least three main contributions. %
First, the intrinsic phase variations that are the goal of the measurement. %
Second, variations of initial parameters from event to event originating from the bunch or the plasma. %
We measure rms variations in bunch length, $\approx$1.6\%, population, $\approx$5\%, and plasma density, $<$0.2\%. %
There may be additional variations in bunch waist size and location, and emittance that we do not monitor for each event. %
The influence of these variations on the phase can in principle be obtained from numerical simulations~\cite{bib:moreira}, though reaching percent level precision is very challenging. %
Third, variations due to the measurement accuracy influenced by the streak camera resolution of the modulation, the limited number of micro-bunches per image, signal noise, and uncertainties in determining the position of the LRS (0.16\,ps).
As a consequence, the measured variations can only be seen as an upper limit for the real phase variations. They are probably dominated by the last two contributions mentioned, mainly by uncertainties originating from the noisy measured modulation profile (see \emph{Supplemental Material}). %

The initial transverse wakefield amplitude (at the plasma entrance) can be calculated as a function of the RIF timings of Fig.~\ref{fig:transitionplot}: $W_{r,RIF}(t=t_{RIF})$. %
The initial proton bunch density ($n_{b0}=1.1\times10^{13}\,\mathrm{cm}^{-3}$) is smaller than the plasma density ($n_{e0}=0.94\times10^{14}\,\mathrm{cm}^{-3}$). % 
We thus use two-dimensional linear plasma wakefield theory~\cite{bib:keinigs} to evaluate this amplitude. %
The modulation period ($\cong$11.5\,ps) is much shorter than the rms bunch duration ($\sigma_{t}$=250\,ps). %
We therefore consider the bunch density constant over one period behind the RIF and thus %
$W_{r,RIF}=2\left(m_ec^2/e\right)\left(n_{b0}(t_{RIF})/n_{e0}\right)\frac{dR}{dr}|_{r=\sigma_{r0}}$. %
The radial dependence of wakefields through the $R(r)$ coefficient~\cite{bib:keinigs} is a function of the transverse bunch profile, considered as Gaussian. %

We plot the amplitude of $W_{r,RIF}$ for each data point %$t_{RIF}$
in Fig.~\ref{fig:transitionplot} (filled red circles). %
The input parameter variations mentioned above cause a maximum statistical uncertainty of 10.1\% on the field calculation, which includes a 15\,ps (0.06$\sigma_t$) rms timing jitter between the proton bunch and the laser pulses (RIF and LRS), all added in quadrature. This uncertainty is indicated by the error bars. %
The plot shows that for the parameters of these experiments, the transition between SMI and SSM occurs between $\sim(2.8\pm0.3)$ and $\sim(4.1\pm0.4)$\,MV/m. % 
The fact that initial wakefield amplitudes of $(2.8\pm0.3)$\,MV/m do not seed the SM process may indicate that the bunch has density irregularities driving initial wakefields with amplitude (between $(2.8\pm0.3)$ and $(4.1\pm0.4)$\,MV/m) much larger than those of the shot noise assumed in \cite{bib:lotovnoise}  driving $<$100\,kV/m fields. We also note here that we interpret the reproducibility of the bunch modulation as also that of the wakefields driven towards the end of the plasma, after saturation of the SM process~\cite{bib:marlene2}. %
The wakefield structure is intrinsically linked to the distribution of the self-modulated proton bunch. %

The phase reproducibility can be further confirmed by similar phase variation measurements at various delays behind the RIF. %
Sets of approximately ten images with delay increments of 50\,ps between each set were acquired at a higher plasma density  $n_{e0}=1.81\times10^{14}\,\mathrm{cm}^{-3}$ and a fixed RIF timing of 125\,ps ($0.5\sigma_{t}$). %
%For these parameters, $W_{r,RIF}=17.3$\,MV/m, which is approximately four times larger than the values for which SSM is observed in Fig.~\ref{fig:transitionplot} for a plasma density only approximately two times larger, thus expectedly placing SM in the SSM regime. %
For these parameters, $W_{r,RIF}=17.3$\,MV/m. As $W_{r,RIF}$ is approximately four times larger, and the density only two times smaller, than the values for which SSM is observed in Fig.~\ref{fig:transitionplot}, SSM is also expected in this case. %
Due to the time overlap between sets, all images can be ``stitched" together using the LRS as described in~\cite{bib:fabianEAAC} (see Fig.~\ref{fig:alongbunch}~(a)). %
It is immediately clear from the figure that micro-bunches of all events align themselves in time/phase and form a coherent modulation of the bunch density over $\sim2\sigma_{t}$ behind the RIF. %
This is only possible when proper seeding is provided (SSM) for each event, relative phase variations between events are small (i.e., all sequences look similar to that of Fig.~\ref{fig:waterfall}~(b)), and the modulation phase is reproducible all along the bunch. %
All features visible in Fig.~\ref{fig:alongbunch}~(a) would wash out if phases were randomly distributed as in Fig.~\ref{fig:waterfall}~(a). %

Figure~\ref{fig:alongbunch}~(b) shows the result of the phase analysis applied to these events. Over the $\sim$2$\sigma_{t}$ measurement range, larger than the delay from the RIF of $\sim1\sigma_{t}$ typically foreseen for external electron injection, the phase variations remain small and in a similar range to those obtained at lower plasma density. %
Variations along the bunch are most likely due to changes in signal that can be seen in Fig.~~\ref{fig:alongbunch}~(a) and on individual images, which affects the accuracy of the phase determination. %
The measured variations remain approximately constant and between 3 to 7\% (of 2$\pi$) all along the bunch. %

\section{Summary}
We presented the results of experimental studies of the SM phase for different timings of the RIF with respect to the proton bunch, measured after the 10\,m-long plasma. %
These results demonstrate that the SM process can be seeded, i.e., the phase of the modulation can be defined by the RIF and reproducible from event to event. %
We observe the transition from phase non-reproducibility and instability (SMI) to seeding and phase reproducibility (SSM) when the transverse wakefield at the RIF exceeds a threshold amplitude, between $(2.8\pm0.3)$ and $(4.1\pm0.4)$\,MV/m for $n_{e0}$=0.94$\times$10$^{14}$\,cm$^{-3}$. %
We show that in the SSM regime variations of the modulation phase along the bunch ($\sim2\sigma_t$) are small, measured at $\le$7\%. % and estimated at $\le$3\% of 2$\pi$. %
We attribute most of these small variations to the measurement accuracy of the modulation phase within single, 73\,ps time windows including only six to nine modulation periods. %
%Studies of the SM phase at various positions along the proton bunch with a fixed seeding/RIF timing confirm the observed phase reproducibility for the entire measurement range of up to 2$\sigma_{t}$ of the proton bunch duration. %
The phase reproducibility also observed at higher plasma density allows for detailed observation of the SM process along the whole bunch with $\sim$ps time resolution %, as shown on
(Fig.~\ref{fig:alongbunch}~(a)). %

Based on these results, one can thus expect that for the studies of electron acceleration during AWAKE Run II~\cite{bib:muggliEAAC}, the wakefields driven by the bunch train in the second plasma, will have a timing/phase also reproducible from event to event since they will be driven by the bunch emerging from the first plasma. %
Phase reproducibility is required for deterministic acceleration of electrons externally injected into the wakefields, with a fixed delay with respect to the seed. %

\section*{Acknowledgements}
This work was supported by the Wolfgang Gentner Programme of the German Federal Ministry of Education and Research (grant no.\ 05E15CHA); in parts by a Leverhulme Trust Research Project Grant RPG-2017-143 and by STFC (AWAKE-UK, Cockcroft Institute core and UCL consolidated grants), United Kingdom; a Deutsche Forschungsgemeinschaft project grant PU 213-6/1 ``Three-dimensional quasi-static simulations of beam self-modulation for plasma wakefield acceleration''; the National Research Foundation of Korea (Nos.\ NRF-2016R1A5A1013277 and NRF-2020R1A2C1010835); the Portuguese FCT---Foundation for Science and Technology, through grants CERN/FIS-TEC/0032/2017, PTDC-FIS-PLA-2940-2014, UID/FIS/50010/2013 and SFRH/IF/01635/2015; NSERC and CNRC for TRIUMF's contribution; the U.S.\ National Science Foundation under grant PHY-1903316; and the Research Council of Norway. M. Wing acknowledges the support of DESY, Hamburg. Support of the Wigner Datacenter Cloud facility through the ``Awakelaser" project is acknowledged.  The work of V. Hafych has been supported by the European Union's Framework Programme for Research and Innovation Horizon 2020 (2014--2020) under the Marie Sklodowska-Curie Grant Agreement No.\ 765710. The AWAKE collaboration acknowledge the SPS team for their excellent proton delivery.

\clearpage
\newpage
\section{Supplemental Material}
\begin{figure}[b]
\centering
\includegraphics[width=\columnwidth]{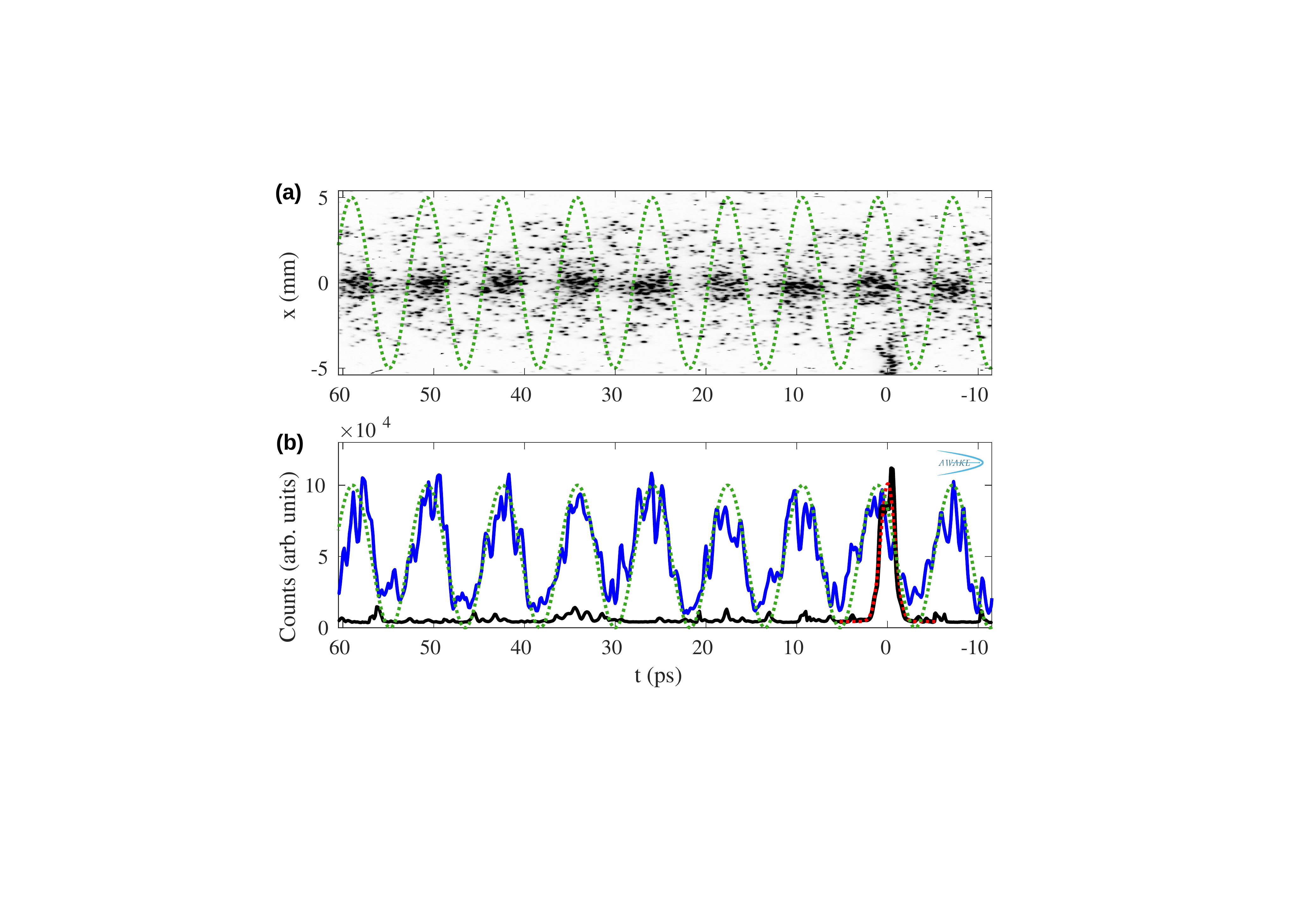}
\caption{(a) Single 73\,ps streak camera image of a self-modulated proton bunch recorded at density $n_{e0}=1.81\times10^{14}\,$cm$^{-3}$. %
(b) Blue line: Bunch modulation profile summing a $\cong\pm$430\,\textmu m-wide region around the bunch axis of (a). Black line: Summed counts over a %a $\cong\pm$1\,mm-wide 
$-5\le x\le -4$\,mm region containing only the LRS. %
Red dotted line: Gaussian fit to determine the position of the LRS. %
Green line, both Figs: cosine of the average DFT frequency used for the analysis and the measured phase value.}
\label{fig:bunchprofile}
\end{figure}

We detail here the analysis that yields the timing/phase variations plotted in Figs~\ref{fig:transitionplot} and \ref{fig:alongbunch}. %
Figure~\ref{fig:bunchprofile}~(a) shows an example of a single time-resolved streak camera image of the self-modulated proton bunch in a 73\,ps window and of the LRS placed 150\,ps (0.6$\sigma_t$) ahead of the bunch center. %
The laser pulse that creates the RIF is 600\,ps (2.4$\sigma_t$) ahead of the bunch center. %
We obtain the  time profile of the bunch density by summing counts of the images in a $\cong\pm$430\,\textmu m-wide region around the bunch axis at the %optical transition radiation
OTR screen (see Fig.~\ref{fig:setup}). %
At this location, the incoming bunch transverse rms size is $\cong$574\,\textmu m. %
The time profile as shown in Fig.~\ref{fig:bunchprofile}~(b) (blue line) consists of 512 amplitude values for the 73\,ps-long window. %

For the modulation frequency and phase determination, we use a discrete Fourier transform (DFT) analysis of the signal. %
The size of the frequency bin of the DFT is given by $\Delta f_{DFT}$=1/73\,ps$\cong$13.7\,GHz. %
We decrease the size of the bin to increase the resolution of the frequency determination by padding the time profile with an array of 50$\times$512 zero amplitudes. %
This procedure brings $\Delta f_{DFT}$ to 13.7\,GHz/50$\cong$0.27\,GHz, %THIS IS 0.3% OF THEVMODULATION FREQUENCY .... IS IT CONSISTENTVWITH THE 0.1% BELOW??
which is on the order of the Rb vapor and thus plasma density measurement accuracy \cite{bib:fabian}. Zero padding is equivalent to including the effect of the 73\,ps-long square function time window of the streak camera image on a longer signal. %
We note here that in signal processing, a Gauss- or Hann-like window function can be used to decrease the convolution effect of the very broad spectrum of the square window $sinc$ function on the signal power spectrum. %
However, with the only six to nine micro-bunches (or DFT periods) in a 73\,ps window of this experiment, such window functions do not improve the quality or precision of the analysis. %
In addition, we only determine the frequency and phase using the DFT signal and not its width or amplitude. Thus we do not use these window functions. %
Further we note that we do not include the data set with $t_{RIF}=0$ (Fig.~\ref{fig:alongbunch}~(a)) in the analysis because images include the non-modulated part of the bunch not in plasma ($t$$<$0\,ps) as well as the onset of the self-modulation.

For a set of images acquired with the same LRS delay \cite{bib:fabianEAAC} with respect to the RIF, we first determine the modulation frequency of the periodic bunch train as the frequency with the largest amplitude in the DFT power spectrum of the padded time profile. %
We restrict the search to a frequency range corresponding to the width of the $sinc$ DFT function (13.7\,GHz in this case)  %
around the plasma frequency expected from the Rb density measured in the vapor source~\cite{bib:karl}: $n_{e0}$=0.94 and 1.81$\times10^{14}$\,cm$^{-3}$ or $\cong$87.1 and $\cong$120.8\,GHz, respectively. %
The small number of micro-bunches visible on the streak camera images and the significant amount of noise lead to variations in the frequency determination of up to 1.8\,GHz (rms) from event to event. %
These variations ($>$ 1.5\%) are larger than those expected from the measurement of Rb vapor density variations ($<$0.2\%) and corresponding plasma frequency variations ($\propto n_{e0}^{1/2}$, $<$0.1\%).
This is also in contradiction to our SM and phase observations at late times and all along the bunch, seen e.g.~in Fig.~\ref{fig:alongbunch}. % 
Indeed, a global frequency variation of 1.8\,GHz rms observed 50\,ps behind the RIF would correspond to a phase shift of 81 \% of 2$\pi$  500\,ps behind the RIF. %
This is clearly much larger than what is observed here. % and in the SSM cases of Fig.~\ref{fig:transitionplot}. %
For the phase analysis, we therefore select for all events the phase value from the frequency bin corresponding to the average DFT frequency of all events.

\begin{figure}[ht]
\centering
\includegraphics[width=\columnwidth]{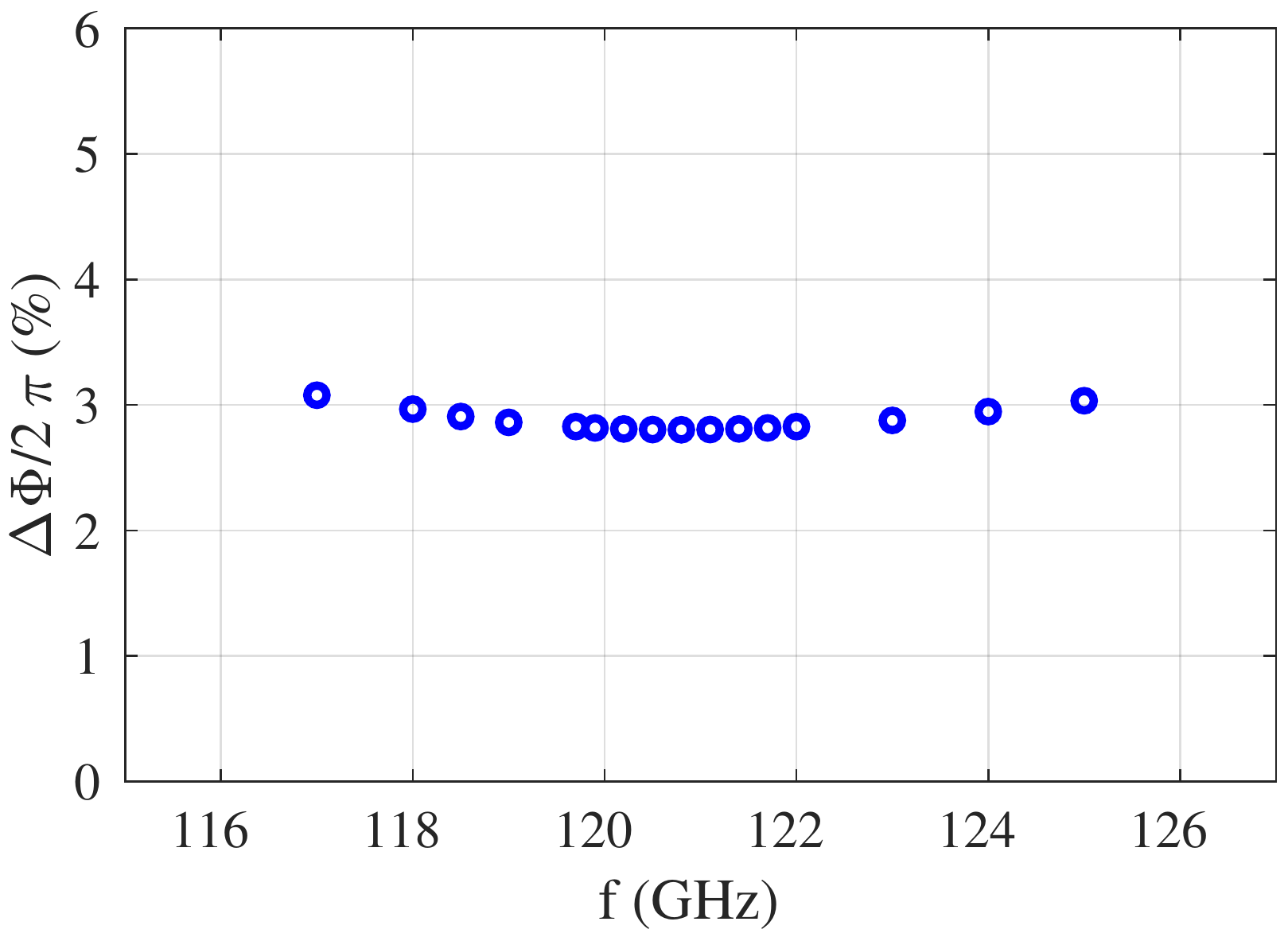}
\caption{Observed phase variation $\Delta\Phi/2\pi$ versus frequency chosen for the phase analysis of a data set recorded at 1.81$\times10^{14}$\,cm$^{-3}$, corresponding to a modulation frequency of 120.8\,GHz.  %
}
\label{fig:rms}
\end{figure}

The variations in the frequency determination do not influence significantly the phase analysis results, as long as these variations are small when compared to the DFT bin width. % 
Figure~\ref{fig:rms} shows the result for the phase variation as a function of the DFT frequency for the data set of Fig.~\ref{fig:alongbunch} at $t$=50\,ps. The figure shows that for all frequencies within $\pm$4\,GHz around the average DFT frequency of 120.8\,GHz, the rms phase variation for these events remains between 2.8 and 3.1\%.
This validates the use of the average DFT frequency for the phase analysis. %
The green dotted lines in Fig.~\ref{fig:bunchprofile} (a) and (b) represent the cosine of the  DFT frequency and the  phase and thus the result of the micro-bunch train determination. %
It shows that the DFT analysis of the bunch time profile with the average modulation frequency and corresponding phases represents the data well. %

We determine the timing of the LRS (-1$\le t\le$+1\,ps and $x\le$-3\,mm in Fig.~\ref{fig:bunchprofile}~(a)) as described in Ref.~\cite{bib:fabianEAAC} by fitting a Gaussian function to its time projection. %
The expected response of the streak camera to signals shorter than its intrinsic time resolution, 120\,fs for the LRS, is $\sim$1\,ps. %
The red dotted line in Fig.~\ref{fig:bunchprofile}~(b) shows the result of the Gaussian fit that represents the LRS. %
Measurements with RIF and LRS laser pulses on the same 73\,ps window show that their time difference is measured with 0.16\,ps precision. %
The mechanical delay line for the LRS has a position accuracy that corresponds to 0.53\,ps (rms). %
This accuracy only influences the time at which phase variation results are plotted in Fig.~\ref{fig:alongbunch}~(b), not their value. %
However, this limited accuracy affects how well the image data sets acquired with different LRS timing ``stitch" together on \ref{fig:alongbunch}~(a). % 

From the timing of the LRS, we determine the time difference and phase between the LRS and the next maximum (micro-bunch) of the periodic function using the average DFT frequency.
For the case of the event of Fig.~\ref{fig:bunchprofile}, the time difference we find is 1.1\,ps, corresponding to a phase difference of 0.8\,rad at the frequency of 120.8\,GHz. %

We repeat this procedure for all events in each series with the same LRS delay with respect to the RIF. %
We characterize the results for each delay set by the rms of the phase distribution as well as by the full range of phases in the distribution, both in \% of 2$\pi$. %
%Since the phase is a modulo 2$\pi$ parameter, the rms of its distribution is only meaningful when it is much smaller than 50\% of 2$\pi$. %
%We therefore quote it only in this case, and quote the range of the distribution for all cases. %

Looking at Fig.~\ref{fig:bunchprofile} it is clear that a significant fraction of the phase variations we measure at the percent level (see main text) could be associated with the time identification of the two signals, at the picosecond level, from single, noisy streak camera images. %
%Due to this noise, the choice of the counts are summed can induce an uncertainty in phase of up to 4\% of 2$\pi$.


\begin{thebibliography}{99}
%1
\bibitem{bib:tajima}T. Tajima, J. M. Dawson, Phys. Rev. Lett. 43, 267 (1979)

\bibitem{bib:chenpwfa}P. Chen {\it et al.}, Phys. Rev. Lett. 54, 693 (1985)

\bibitem{bib:chenfoc}P. Chen {\it et al.}, IEEE Trans. Plasma Sci. 15(2), 218 (1987), P. Chen {\it et al.}, Phys. Rev. D 40, 923 (1989)

\bibitem{bib:blumenfeld}I. Blumenfeld {\it et al.}, Nature 445, 741-744 (2007)

\bibitem{bib:mike}M. Litos {\it et al.}%E. Adli, W. An,	C. I. Clarke, C. E. Clayton, S. Corde, J. P. Delahaye, R. J. England, A. S. Fisher, J. Frederico, S. Gessner, S. Z. Green, M. J. Hogan, C. Joshi, W. Lu, K. A. Marsh, W. B. Mori, P. Muggli, N. Vafaei-Najafabadi, D. Walz, G. White, Z. Wu, V. Yakimenko, G. Yocky
, Nature 515, 92 (2014)
%5
\bibitem{bib:benedeti}A. J. Gonsalves {\it et al.}%, K. Nakamura, J. Daniels, C. Benedetti, C. Pieronek, T. C. H. de Raadt, S. Steinke, J. H. Bin, S. S. Bulanov, J. van Tilborg, C. G. R. Geddes, C. B. Schroeder, Cs. Tóth, E. Esarey, K. Swanson, L. Fan-Chiang, G. Bagdasarov, N. Bobrova, V. Gasilov, G. Korn, P. Sasorov, and W. P. Leemans
, Phys. Rev. Lett. 122, 084801 (2019)

\bibitem{bib:wavebreaking}J. M. Dawson, Phys. Rev. 113, 383 (1959)

\bibitem{bib:cst}Constants have usual meaning: $e$   electron charge, $\varepsilon_0$  vacuum permittivity, $m_e$  electron mass, $c$  speed of light in vacuum

%\bibitem{bib:roswell}R. Lee, M. Lampe, Phys. Rev. Lett. 31(23), 1390 (1973)
\bibitem{bib:su}J. J. Su {\it et al.},  IEEE Transactions on Plasma Science 15, 192–198 (1987)
%10
\bibitem{bib:laserSM}C. J. McKinstrie, Physics of Fluids B: Plasma Physics 4, 2626 (1992)

\bibitem{bib:protonSM}N. Kumar {\it et al.}, Phys. Rev. Lett. 104, 255003 (2010)

%\bibitem{bib:raman}E. Esarey, C. B. Schroeder, and W. P. Leemans, Rev. Mod. Phys. 81, 1229 (2009)

%\bibitem{bib:rosenzweig}J. B. Rosenzweig {\it et al.}, Phys. Rev. Lett. 61, 98 (1988)

%\bibitem{bib:caldwellTeV}A. Caldwell {\it et al.}, Phys. Plasmas 18, 103101 (2011)
%15
%\bibitem{bib:wing}M. Wing, Philosophical Transactions of the Royal Society A: Mathematical, Physical and Engineering Sciences, 377, 20180185 (2019).

\bibitem{bib:awake}E. Gschwendtner {\it et al.},  Nucl. Instr. and Meth. in Phys. Res. A 829, 76 (2016)

\bibitem{bib:muggli}P. Muggli {\it et al.} (AWAKE Collaboration), Plasma Physics and Controlled Fusion, 60(1) 014046 (2018)

\bibitem{bib:karl}E. Adli {\it et al.} (AWAKE Collaboration), Phys. Rev. Lett. 122, 054802 (2019)

\bibitem{bib:marlene}M. Turner {\it et al.} (AWAKE Collaboration), Phys. Rev. Lett. 122, 054801 (2019)
%16
\bibitem{bib:marlene2}M. Turner, P. Muggli {\it et al.}, (AWAKE Collaboration), Phys. Rev. Accel. Beams 23, 081302 (2020)

\bibitem{bib:awakeaccel}E. Adli {\it et al.} (AWAKE Collaboration), Nature 561, 363 (2018), E. Gschwendtner, M. Turner  {\it et al.} (AWAKE Collaboration), Phil. Trans. R. Soc. A 377, 20180418 (2019)

\bibitem{bib:veronica}V. K. Berglyd Olsen, E. Adli, P. Muggli, Phys. Rev. Accel. Beams 21, 011301 (2018)

\bibitem{bib:leblanc}S. P. Le Blanc {\it et al.},
Phys. Rev. Lett. 77, 5381 (1996)
%20
\bibitem{bib:fang}Y. Fang {\it et al.}, Phys. Rev. Lett. 112, 045001 (2014)

\bibitem{bib:moreira}M. Moreira, J. Vieira, P. Muggli, Phys. Rev. Accel. Beams 22, 031301 (2019)

\bibitem{bib:gennady}G. Plyushchev {\it et al.}, J. Phys. D: Appl. Phys. 51, 025203 (2017)

\bibitem{bib:erdem}E. \"Oz {\it et al.}, Nucl. Instr. and Meth. in Phys. Res. A 740, 197 (2014)

\bibitem{bib:fabian}F. Batsch {\it et al.}, Nucl. Instr. and Meth. in Phys. Res. A 909, 359 (2018)

\bibitem{bib:annamaria}A.-M. Bachmann, P. Muggli, J. Phys.: Conf. Ser. 1596, 012005 (2020)
%26
\bibitem{bib:fabianEAAC}F. Batsch, J. Phys.: Conf. Ser. 1596, 012006 (2020)

\bibitem{bib:gessner}S. Gessner {\it et al.} (AWAKE Collaboration),  submitted (2020), arXiv:2006.09991

\bibitem{bib:lotovnoise}K. Lotov {\it et al.}, Phys. Rev. ST Accel. Beams 16, 041301 (2013)

\bibitem{bib:muggliEAAC}P. Muggli {\it et al.},  J. Phys.: Conf. Ser. 1596, 012008 (2020)
%30
\bibitem{bib:keinigs}R. Keinigs, M. E. Jones, Phys. Fluids 30, 252 (1987)

%\bibitem{bib:mathias}M. Huether {\it et al.}, in preparation

%\bibitem{bib:laserjitter}S. Schultz {\it et al.}, Nature Communications 6, 5938 (2015), Dohyeon Kwon, Chan-Gi Jeon, Dohyun Kim, Igju Jeon, and Jungwon Kim, Optics Letters 45(11), 3155 (2020)

\end{thebibliography}
\end{document}